\documentclass[aps,prl,reprint,a4paper,showkeys,showpacs]{revtex4-1}
\usepackage{amsmath}
\usepackage{amssymb}
\usepackage{amsfonts}
\usepackage{graphicx}

\begin{document}

\title{Strong Chaos of Fast Scrambling Yields Order:
Emergence of Decoupled Quantum Information Capsules}
\ \\
\author{Masahiro Hotta}
\author{Koji Yamaguchi}
\affiliation{Graduate School of Science, Tohoku University,\\ Sendai, 980-8578, Japan}

\begin{abstract}
The information loss problem in black hole evaporation is one of fundamental issues.
Its resolution requires more profound understanding of information storage mechanism in quantum systems.
In this Letter,  we argue that when multiple unknown parameters are stored in large entangled qudits,
strong chaos generated by fast scrambling in high temperature limit yields
an ordered information storage structure with decoupled quantum information capsules (QICs).
A rotational isometry emerges in the quantum Fisher information metric.
The isometry is expected to be observed in future experiments on cold atoms in a pure entangled state.
We provide a QIC speculation of black hole evaporation.
\end{abstract}
\maketitle

 \textit{Introduction}.--- Fast scrambling is supposed to be crucial for
understanding of the information loss problem in black hole evaporation
\cite{HP,SS}. If unitarity is maintained in quantum gravity, black hole
systems store information of infalling matters without destroying it. The fast
scrambling inside the horizons is rapid information scrambling described by
random unitary operations. Hayden and Preskill \cite{HP} argued that such information mixing generates black hole mirrors, and forces the stored
information to be retrieved outside in a much shorter duration than the black
hole lifetime. Besides, many other interesting properties of black holes as
information storages have been reported including no-hiding theorem \cite{BP}
and no-masking theorem \cite{P}. As well as quantum black holes, general
quantum macroscopic systems in entangled pure states are capable of storing
information of unknown parameters. 
Investigating general structures of information storage for general many-body
systems sheds light on the black hole information loss problem. In recent
development of quantum technology, the systems in pure entangled states are
becoming available experimentally \cite{RD,SKA}. Especially, entanglement detection was
achieved using cold atoms in optical lattice \cite{ED}. Similarly, quantum
Fisher information metric of a pure state $|\Psi\left( \theta_{1}
,\cdots,\theta_{n}\right)  \rangle_{1\cdots N}$ depending on multiple
parameters $\left(  \theta_{1},\cdots,\theta_{n}\right)$ may be expected to
be directly measured in future experiments. The metric is defined as
\begin{align}
&\left[  g_{jj^{\prime}}\right]\nonumber  \\
&=\left[  \langle\Psi\left(  \theta_{1}
,\cdots,\theta_{n}\right)  |\frac{1}{2}\left(  L_{j}L_{j^{\prime}
}+L_{j^{\prime}}L_{j}\right)  |\Psi\left(  \theta_{1},\cdots,\theta
_{n}\right)  \rangle\right]\nonumber,
\end{align}
where symmetric logarithmic derivative operators $L_{j}$ are given by
\begin{align}
&L_{j}=
|\partial_{\theta_{j}}\Psi\left(  \theta_{1},\cdots,\theta_{n}\right)
\rangle\langle\Psi\left(  \theta_{1},\cdots,\theta_{n}\right)  |\nonumber\\
&\quad \quad+|\Psi\left(
\theta_{1},\cdots,\theta_{n}\right)  \rangle\langle\partial_{\theta_{j}}
\Psi\left(  \theta_{1},\cdots,\theta_{n}\right) |. 
\end{align}
It is pointed out that the entire family of quantum Fisher information can be
determined from linear-response theory through generalized covariances
\cite{SU}. By observing the covariances, the quantities may be determined by
experiments. The quantum Fisher information metric elucidates
information storage structures for quantum systems.

As an information storage of an unknown continuous parameter $\theta$, let us
consider $N$ qudits in a pure state $|\Psi\rangle_{12\cdots N}$. A write
operation $W(\theta)_{1\cdots N}=\exp\left(  i\theta\sigma\right)
\otimes I_{2}\otimes\cdots\otimes I_{N}$ of $\theta$ is performed to the
first qudit as its target subsystem, where $\sigma$ is an element of $SU(d)$
algebra. It is worth stressing that for the fixed system, the fixed state
$|\Psi\rangle_{12\cdots N}$, and the fixed write operation $W(\theta)_{1\cdots N}  $, there exist three different pictures for the
information storage of $\theta$ in quantum mechanics. One of them is the standard
one, and is referred to as Schmidt partner picture in this Letter. Consider the Schmidt decomposition of $|\Psi\rangle_{12\cdots N}$ between the first
qudit and other qudits
\[
|\Psi\rangle_{12\cdots N}=\sum_{k=1}^{d}\sqrt{p_{k}}|u_{k}\rangle_{1}
|v_{k}\rangle_{2\cdots N},
\]
where $p_{k}$ is a probability distribution, and $|u_{k}\rangle_{1}$ are
orthonormal basis vectors of the first qudit, and $|v_{k}\rangle_{2\cdots N} $
are orthonormal vectors for the remaining system. Using a local unitary
operation $I_{1}\otimes\tilde{U}_{2\cdots N}$, the state can be written as
\[
|\Psi\rangle_{12\cdots N}=\left(  I_{1}\otimes\tilde{U}_{2\cdots N}\right)
\left(  \sum_{k=1}^{d}\sqrt{p_{k}}|u_{k}\rangle_{1}|\varphi_{k}\rangle
_{2}\right)  |\chi\rangle_{3\cdots N},
\]
where $|\varphi_{k}\rangle_{2}$ are orthonormal basis vectors of the second
qudit, and $|\chi\rangle_{3\cdots N}$ is a decoupled pure state of $N-2$
qudits. Then its corresponding pure composite state of the target $A~$(the
first qudit) and its Schmidt partner $B$\ is given by $|\Psi_{S}\rangle
_{AB}=\sum_{k}\sqrt{p_{k}}|u_{k}\rangle_{A}|\varphi_{k}\rangle_{B}.$ After the
write operation, we get the target-partner state for
$\theta$ as $|\Psi_{S}\left(  \theta\right)  \rangle_{AB}$ $=\sum_{k}
\sqrt{p_{k}}\left(  \exp\left(  i\theta\sigma\right)  |u_{k}\rangle
_{A}\right)  |\varphi_{k}\rangle_{B}$. This is the first picture of quantum
information storage of $\theta$. The second picture is
referred to as generalized partner picture, which was proposed in
\cite{TYH2,TYH}. Let us introduce a
picture-change unitary operator $V_{AB}\left(  \beta\right) $ with a real
continuous parameter $\beta$, which satisfies $V_{AB}\left(  \beta=0\right) =I_A\otimes I_B$ and commutes with the write operation:
\begin{equation}
\left[  V_{AB}\left(  \beta\right) ,  \exp\left(
i\theta\sigma\right)  \otimes I_{B}\right]  =0\label{1}.
\end{equation}
Performing $V\left(  \beta\right)  _{AB}$ to $|\Psi_{S}
\rangle_{AB}$ generates other information storage states given by
$|\tilde{\Psi}\left(  \theta,\beta\right)  \rangle_{AB}=\left(  \exp\left(
i\theta\sigma\right)  \otimes I_{B}\right)  V\left(  \beta\right)  _{AB}
|\Psi_{S}\rangle_{AB}=V\left(  \beta\right)  _{AB}|\Psi_{S}\left(
\theta\right)  \rangle_{AB}$. 
In this
case, the entanglement entropy $S_{EE}$ between target $A$ and its partner $B$ depends on $\beta$.
If $S_{EE}\neq0$ after performing the unitary $V(\beta)$, the new picture is referred to as generalized partner picture.
If $S_{EE}$ vanishes for a certain
value $\beta=\beta_{\ast}$, $V(\beta_*)|\Psi_S(\theta)\rangle_{AB}=|\tilde{\Psi}(\theta)\rangle_A|\psi\rangle_B$ holds. 
Then, $A$ after
the picture change operation is referred to as quantum information capsule
(QIC) for $\theta$ \cite{YWH}. This is QIC picture for information storage of
$\theta$. Surprisingly, even under the constraint in eq. (\ref{1}),
there exists a value $\beta_{\ast}$ for an arbitrary $|\Psi
\rangle_{12\cdots N}$. Hence the QIC picture can always be adopted. As
opposed to the QIC existence, it is not always possible to find a value of
$\beta$\ such that $S_{EE}$ takes its maximum value ($S_{EE}=\ln d$). Only
specific $|\Psi\rangle_{12\cdots N}$ allows to attain the maximum. This
emphasizes the non-triviality of QIC existence again. In the QIC picture, only
a single qudit is required, though two qudits are needed in both Schmidt
partner picture and generalized partner picture. Hence, in dynamical
situations, QIC pictures make analysis of information flow of multiple
parameters much simpler.

In this Letter, we show that strong chaos generated by fast scrambling yields
a counterintuitive phenomenon. Ordered structure of information storage emerges for $n$ parameters
$\theta_{j}$ ($j=1,\cdots,n$) in double limits of large $N$ and high
temperature $T$ for pure thermal states $|\Psi\left(  T\right)  \rangle
_{1\cdots N}$ of $N$ qudits. The pure thermal states $|\Psi\left(  T\right)
\rangle_{1\cdots N}$ are defined to exhibit the same thermodynamical behaviors for macroscopic observables as the Gibbs states at temperature $T$. Their
construction is given as follows. Let us consider a free Hamiltonian $H_{j}$ for each qudit. 
The eigenstates of total Hamiltonian $H=\sum_{j=1}^{N}H_{j}$ are denoted
by $|E\rangle_{1\cdots N}$ with corresponding eigenvalue $E$. For a fixed
value of total energy $E_{tot}$ and small energy difference $\delta E$, the microcanonical energy shell (MES) is defined in a standard way \cite{L,SJ,SA,S,S1} as the Hilbert space spanned by energy eigenstates $|E\rangle_{1\cdots N}$ satisfying $E\in[E_{tot}-\delta E,E_{tot}]$.
Taking MES ensemble average in the uniform
distribution of MES subHilbert space, it is proven that the reduced state of small subsystems for a typical state of the total system in MES becomes a Gibbs
state with temperature $T$ \cite{SM}. In high temperature
limit, typical states in strong chaos are generated by random unitary
operators $U$ of the total system as $|\Psi\left(  T\rightarrow\infty\right)  \rangle_{1\cdots N}=U|0\rangle
_{1\cdots N}$,
where $|0\rangle_{1\cdots N}$ is an arbitrarily fixed initial state.
Note that the high temperature limit is equivalent to zero Hamiltonian limit
in which all the eigenstates of $H$ are degenerated. Such a system is experimentally available in spinor Bose gases \cite{SBG1,SBG2} and Fermi gases in an optical lattice \cite{FG1,FG2}. For example, the
effective Hamiltonian of cold atoms is given by $H=0$. 

The main setup of our problem is the following. 
First, let us perform the write operation $W\left(  \theta_{1}\right)  $ for
$\theta_{1}$ to a typical state $|\Psi\rangle_{12\cdots N}$ belonging to MES at high temperature.
After that, a scrambling operation $U_{1}$ among MES is performed. By repeating these processes $n$
times, we get the following encoded state:
\begin{align}
&|\Psi\left(  \theta_{1},\cdots,\theta_{n}\right)  \rangle_{1\cdots N}\nonumber\\
&=U_{n}W\left(  \theta_{n}\right)  U_{n-1}W\left(  \theta_{n-1}\right)  \cdots
U_{1}W\left(  \theta_{1}\right)  |\Psi\rangle_{12\cdots N}.\label{5}
\end{align}
In general, QICs of the parameters are extremely tangled with each other. The
quantum state has a very complicated structure of information storage. A tangled $\theta_{j}$-QIC is constructed by
\begin{align}
&|\Psi\left(  \theta_{1},\cdots,\theta_{n}\right)  \rangle_{1\cdots N}\nonumber\\
&=S_{code}\left(  \theta_{j+1},\cdots,\theta_{n}\right)
|\psi_{\theta_{1},\cdots,\theta_{j-1}}\left(  \theta_{j}\right)  \rangle
_{1}|\Upsilon\left(  \theta_{1},\cdots,\theta_{j-1}\right)
\rangle_{2\cdots N},\label{3}
\end{align}
where $S_{code}\left(  \theta_{j+1},\cdots,\theta_{n}\right)$ is a unitary operation depending on $\theta
_{1},\cdots,\theta_{j-1}$, $|\psi_{\theta_{1},\cdots,\theta_{j-1}}\left(
\theta_{j}\right)  \rangle_{1}$ is a $\theta_{j}$-QIC state for the
first qudit, and $|\Upsilon\left(  \theta_{1},\cdots,\theta_{j-1}\right)
\rangle_{2\cdots N}$ is a quantum state of the other qudits depending on
$\theta_{1},\cdots,\theta_{j-1}$. 
Interestingly, by taking the large $N$ limit and high temperature limit, the
structure is drastically simplified. Decoupled QICs emerge in high precision,
and each QIC confines the information of each $\theta_{j}$ as
\begin{align}
&|\Psi\left(  \theta_{1},\cdots,\theta_{n}\right)  \rangle_{1\cdots N}\nonumber\\
&=U_{code}|\varphi\left(  \theta_{1}\right)  \rangle_{1}|\varphi\left(  \theta
_{2}\right)  \rangle_{2}\cdots|\varphi\left(  \theta_{n}\right)  \rangle_{n}
|\Psi\rangle_{n+1\cdots N},\label{2}
\end{align}
where $U_{code}$ is a unitary operation independent of $\theta_{1}
,\cdots,\theta_{n}$. Its corresponding quantum Fisher information metric is a
constant metric proportional to unit matrix as $\left[  g_{jj^{\prime}}\right]
= F\delta_{jj^{\prime}} $ with $\partial_{\theta_{j}}F=0$. Thus, a rotational isometry emerges in the parameter space. For an arbitrary
rotation matrix $R=[R_{jj^{\prime}}]\,$, the redefinition of parameters as
$\theta_{j}^{\prime}=\sum_{j^{\prime}}R_{jj^{\prime}}\theta_{j}$ in
$|\Psi\left(  \theta_{1},\cdots,\theta_{n}\right)  \rangle_{1\cdots N}$
provides the same quantum Fisher information metric $\left[  F\delta
_{jj^{\prime}}\right]  $. At low temperature, the isometry is broken.
Our results may be verified in future experiments on cold atoms in pure states.

\begin{figure}[tbp]
\includegraphics[width=4.0cm]{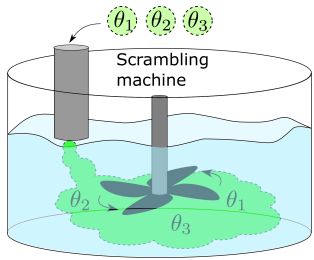}
\caption{Schematic picture for a water solution analogy of tangled QICs at low temperature.}\label{figure_lt}
\end{figure}

\begin{figure}[tbp]
\includegraphics[width=4.0cm]{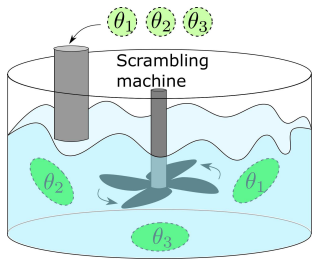}
\caption{Schematic picture for a water solution analogy of decoupled QICs at high temperature.}\label{figure_ht}
\end{figure}

Figure \ref{figure_lt} and \ref{figure_ht} are schematical figures which describe a water solution analogy
of the above QIC picture. 
In Figure \ref{figure_lt}, water at low temperature is in a box
and is stirred up by a scrambling machine. Droplets of different density
$\theta_{1},\theta_{2},\theta_{3}$ of a chemical are injected to the box. The
droplets make a quite tangled cloud in the water carrying the information of
$\theta_{1},\theta_{2},\theta_{3}$, corresponding to the state in eq.
(\ref{3}). In Figure \ref{figure_ht}, a case with high temperature water in strong chaos is
depicted. The decoupled QICs emerge and confine the information separately.
This highlights the counterintuitive feature of the ordered QIC structure in
eq. (\ref{2}) generated by strong chaos in high temperature regime. 

In this Letter, we adopt
the natural units: $c=\hbar=k_{B}=1$.

\bigskip

\textit{Random Unitary Operations.}--- In order to describe strong chaos of $N$
qudits in the double limits of large $N$ and high $T$, $U(d^{N})$-group random
unitary operators are introduced. The error for using the operators with
respect to finite temperature is $O\left(  \Delta E/T\right)  $ where $\Delta
E$ is difference between the maximum and minimum of eigenvalues of $H$.
Suppose that $m$ state vectors $|\lambda\rangle_{1\cdots N}$ with
$\lambda=1,\cdots,m\,$ satisfies $\langle\lambda|\lambda^{\prime}
\rangle=\delta_{\lambda\lambda^{\prime}}$. Dividing the system into the first
qudit and the other $N-1$ qudits, let us introduce an orthonormal basis $\left\{
|a\rangle_{1}|b\rangle_{2\cdots N}|a=1,\cdots,d,b=1,\cdots,d^{N-1}\right\}  $.
Then elements of random unitary operator $U$ are given by $U_{ab,\lambda
}=\langle a|\langle b|U|\lambda\rangle$, and the ensemble averages satisfy the
following relations \cite{SM}:
\begin{equation}
\overline{U_{a_{1}b_{1},\lambda_{1}}U_{a_{2}b_{2},\lambda_{2}}^{\ast}}=\frac
{1}{d^{N}}\delta_{a_{1}a_{2}}\delta_{b_{1}b_{2}}\delta_{\lambda_{1}\lambda
_{2}},\label{6}
\end{equation}

\begin{align}
&  \overline{U_{a_{1}b_{1},\lambda_{1}}U_{a_{2}b_{2},\lambda_{2}}^{\ast}
U_{a_{3}b_{3},\lambda_{3}}U_{a_{4}b_{4},\lambda_{4}}^{\ast}}\nonumber\\
&  =\frac{1}{d^{2N}-1}\left[
\begin{array}
[c]{c}
\delta_{a_{1}a_{2}}\delta_{a_{3}a_{4}}\delta_{b_{1}b_{2}}\delta_{b_{3}b_{4}
}\delta_{\lambda_{1}\lambda_{2}}\delta_{\lambda_{3}\lambda_{4}}\\
+\delta_{a_{1}a_{4}}\delta_{a_{3}a_{2}}\delta_{b_{1}b_{4}}\delta_{b_{3}b_{2}
}\delta_{\lambda_{1}\lambda_{4}}\delta_{\lambda_{3}\lambda_{2}}
\end{array}
\right] \nonumber\\
&  -\frac{1}{d^{N}\left(  d^{2N}-1\right)  }\left[
\begin{array}
[c]{c}
\delta_{a_{1}a_{4}}\delta_{a_{3}a_{2}}\delta_{b_{1}b_{4}}\delta_{b_{3}b_{2}
}\delta_{\lambda_{1}\lambda_{2}}\delta_{\lambda_{3}\lambda_{4}}\\
+\delta_{a_{1}a_{2}}\delta_{a_{3}a_{4}}\delta_{b_{1}b_{2}}\delta_{b_{3}b_{4}
}\delta_{\lambda_{1}\lambda_{4}}\delta_{\lambda_{3}\lambda_{2}}
\end{array}
\right]  .\label{7}
\end{align}
For the reduced state $\rho_{2\cdots N}^{(\lambda)}$ $=\operatorname*{Tr}
_{1}\left[  U|\lambda\rangle_{1\cdots N}\langle\lambda|_{1\cdots N}U^{\dag
}\right]  $ of $N-1$ qudits, the following relation holds:
\begin{equation}
\operatorname*{Tr}_{2\cdots N}\left[  \rho_{2\cdots N}^{(\lambda)}
\rho_{2\cdots N}^{(\lambda^{\prime})}\right]  =
{\displaystyle\sum\limits_{a,a^{\prime}=1}^{d}}
{\displaystyle\sum\limits_{b,b^{\prime}=1}^{d^{N-1}}}
U_{ab,\lambda}U_{ab^{\prime},\lambda}^{\ast}U_{a^{\prime}b^{\prime},
\lambda^{\prime}}U_{a^{\prime}b,\lambda^{\prime}}^{\ast}.
\end{equation}
The Schmidt decomposition of $|\Psi^{(\lambda)}\rangle_{1\cdots N}
=U|\lambda\rangle_{1\cdots N}$ is given by

\begin{equation}
|\Psi^{(\lambda)}\rangle_{1\cdots N}=
{\displaystyle\sum\limits_{a=1}^{d}}
\sqrt{p(a,\lambda)}|u(a,\lambda)\rangle_{1}|\psi(a,\lambda)\rangle_{2\cdots
N},\label{11}
\end{equation}
where $p(a,\lambda)\,$\ is a $\lambda\,$-dependent probability distribution,
$|u(a,\lambda)\rangle_{1}$ are orthonormal basis vectors of the first qudit,
and $|\psi(a,\lambda)\rangle_{2\cdots N}$ is $d$ orthonormal state vectors of
the other qudits. By substituting the decomposition, the following nonnegative
quantity is obtained:
\begin{align}
&\operatorname*{Tr}_{2\cdots N}\left[  \rho_{2\cdots N}^{(\lambda)}
\rho_{2\cdots N}^{(\lambda^{\prime})}\right]  \nonumber\\
&=
{\displaystyle\sum\limits_{a=1}^{d^{N-1}}}
{\displaystyle\sum\limits_{a^{\prime}=1}^{d^{N-1}}}
p(a,\lambda)p(a^{\prime},\lambda^{\prime})\left\vert \langle\psi(a^{\prime
},\lambda^{\prime})|\psi(a,\lambda)\rangle\right\vert ^{2}\geq0.\label{9}
\end{align}
By taking $\lambda^{\prime}=\lambda$, we get $
\sum\limits_{a=1}^{d}
p(a,\lambda)^{2}=\operatorname*{Tr}_{B}\left[  \rho_{2\cdots N}^{(\lambda
)}\rho_{2\cdots N}^{(\lambda)}\right]  $. From eq. (\ref{7}), the ensemble
average $\operatorname*{Tr}_{1}\left[  \overline{\rho_{2\cdots N}^{(\lambda
)}\rho_{2\cdots N}^{(\lambda)}}\right]  $ is calculated for large $N$ as
\begin{align*}
\operatorname*{Tr}_{1}\left[  \overline{\rho_{2\cdots N}^{(\lambda)}
\rho_{2\cdots N}^{(\lambda)}}\right] & =\frac{1}{d^{2N}-1}\left[  d^{2}d^{N-1}+dd^{2(N-1)}\right]  \\
&\quad-\frac{1}
{d^{N}\left(  d^{2N}-1\right)  }\left[  d^{2}d^{N-1}+dd^{2(N-1)}\right] \\
& =\frac{1}{d}\left(1+O\left(  d^{-(N-2)}\right)  \right).
\end{align*}
Note that $1/d$ is the maximum value of $
\sum\limits_{a=1}^{d}
p(a,\lambda)^{2}$, which is attained at $p(a,\lambda)=1/d$. This yields essentially the same result of the Page theorem \cite{Lubkin, Seth_Pagels,
Page_sub} such that
\begin{equation}
p(a,\lambda)=1/d+O\left(  d^{-(N-3)}\right)  .\label{10}
\end{equation}
This implies that $|\Psi^{(\lambda)}\rangle_{1\cdots N}$ typically becomes a maximally
entangled state in large $N$ limit.

By taking the ensemble average of $\operatorname*{Tr}_{2\cdots N}\left[
\rho_{2\cdots N}^{(\lambda)}\rho_{2\cdots N}^{(\lambda^{\prime})}\right]  $,
the following identity is obtained:
\begin{align}
& {\displaystyle\sum\limits_{a=1}^{d^{N-1}}}
{\displaystyle\sum\limits_{a^{\prime}=1}^{d^{N-1}}}
\overline{p(a,\lambda)p(a^{\prime},\lambda^{\prime})\left\vert \langle
\psi(a^{\prime},\lambda^{\prime})|\psi(a,\lambda)\rangle\right\vert ^{2}}\nonumber\\
&=
{\displaystyle\sum\limits_{a=1}^{d}}
{\displaystyle\sum\limits_{a^{\prime}=1}^{d}}
{\displaystyle\sum\limits_{b=1}^{d^{N-1}}}
{\displaystyle\sum\limits_{b^{\prime}=1}^{d^{N-1}}}
\overline{U_{ab\lambda}U_{ab^{\prime}\lambda}^{\ast}U_{a^{\prime}b^{\prime
}\lambda^{\prime}}U_{a^{\prime}b\lambda^{\prime}}^{\ast}}.
\end{align}
Substitution of eq. (\ref{7}) and eq. (\ref{10}) into the above equation yields
\[
\frac{1}{d^{2}}
{\displaystyle\sum\limits_{a=1}^{d^{N-1}}}
{\displaystyle\sum\limits_{a^{\prime}=1}^{d^{N-1}}}
\overline{\left\vert \langle\psi(a^{\prime},\lambda^{\prime})|\psi
(a,\lambda)\rangle\right\vert ^{2}}=O\left(  d^{-(N-1)}\right)
\]
for $\lambda^{\prime}\neq\lambda$. Therefore we obtain a upper bound of
$\left\vert \langle\psi(a^{\prime},\lambda^{\prime}\neq\lambda)|\psi
(a,\lambda)\rangle\right\vert $ as
\begin{equation}
\left\vert \langle\psi(a^{\prime},\lambda^{\prime})|\psi(a,\lambda
)\rangle\right\vert =O\left(  d^{-(N-3)/2}\right)  \to 0 \quad (N\to\infty).\label{15}
\end{equation}
This concludes that $|\psi(a,\lambda)\rangle_{2\cdots N}$ in the Schmidt
decomposition in eq. (\ref{11}) is typically orthogonal to another $|\psi(a,\lambda
^{\prime}\neq\lambda)\rangle_{2\cdots N}$ in large $N$ limit. These results
derive decoupled QICs.

\textit{Strong Chaos Yields Decoupled QICs.}--- Let us take independent random
unitary operators which satisfy the above relations in eqs (\ref{6}) and
(\ref{7}) for $U_{k}$ ($k=1,\cdots,n$) in eq. (\ref{5}). The spectral
decomposition of the write operator is given as $W\left(  \theta\right)
=\sum_{s_{1}=1}^{d}\exp\left(  iw_{s_{1}}\theta\right)  |s_{1}\rangle\langle
s_{1}|\otimes I_{2\cdots N}$, where $\sigma=\sum_{s=1}^d w_{s}|s\rangle\langle s|$ is the eigenvalue decomposition. Let us take the initial typical state as an almost maximally entangled state
\[
|\Psi\rangle_{12\cdots N}=\frac{1}{\sqrt{d}}
{\displaystyle\sum\limits_{s_{1}=1}^{d}}
|s_{1}\rangle_{1}|\psi(s_{1})\rangle_{2\cdots N} +O\left(  d^{-(N-3)/2}\right),
\]
which is generated by a random unitary operator $U_0$ acting on $|0\rangle_{1\cdots N}\equiv|0\rangle_1\cdots|0\rangle_N$.
Performing $W(\theta_1)$ and a random unitary $U_1$ to the initial state provides the following equation:
\begin{align*}
& U_{1}W\left(  \theta_{1}\right)  |\Psi\rangle_{12\cdots N}\\
&=\frac{1}{\sqrt{d}}
{\displaystyle\sum\limits_{s_{1}=1}^{d}}
\exp\left(  iw_{s_{1}}\theta\right)  U_{1}|s_{1}\rangle_{1}|\psi(s_{1}
)\rangle_{2\cdots N}+O\left(  d^{-(N-3)/2}\right)  .
\end{align*}
Let $|\Psi^{(s_{1})}\rangle_{1\cdots N}$ denote $|s_{1}\rangle_{1}|\psi
(s_{1})\rangle_{2\cdots N}$. Since $\langle\Psi^{(s_{1})}|\Psi^{(s_{1}
^{\prime})}\rangle=\delta_{s_{1}s_{1}^{\prime}}\,$holds, $U_{1}|\Psi^{(s_{1}
)}\rangle_{1\cdots N}$ ($s_{1}=1,\cdots,d$) are orthonormal to each other. 
$U_{1}|\Psi^{(s_{1})}\rangle_{1\cdots N}$ is typically a maximally entangled state 
\[
U_{1}|\Psi^{(s_{1})}\rangle_{1\cdots N}=\frac{1}{\sqrt{d}}
{\displaystyle\sum\limits_{s_{2}=1}^{d}}
|s_{2}\rangle_{1}|\psi\left(  s_{1}s_{2}\right)  \rangle_{2\cdots N}.
\]
Note that $\langle\psi\left(  s_{1}s_{2}\right)  |\psi\left(  s_{1}^{\prime}
s_{2}^{\prime}\right)  \rangle=\delta_{s_{1}s_{1}^{\prime}}\delta_{s_{2}
s_{2}^{\prime}}$ is guaranteed from eq. (\ref{15}). The post-scrambling state
is given by
\begin{align}
& U_{1}W\left(  \theta_{1}\right)  |\Psi\rangle_{12\cdots N}\nonumber\\
&=\left(  \frac
{1}{\sqrt{d}}\right)  ^{2}
{\displaystyle\sum\limits_{s_{1}=1}^{d}}
{\displaystyle\sum\limits_{s_{2}=1}^{d}}
\exp\left(  iw_{s_{1}}\theta\right)  |s_{2}\rangle_{1}|\psi\left(  s_{1}
s_{2}\right)  \rangle_{2\cdots N}\nonumber\\
&\quad+O\left(  d^{-(N-3)/2}\right)  \label{eq_one-parameter}.
\end{align}
Repeating the procedure $n$ times generates the encoded state as follows:
\begin{align}
&  |\Psi\left(  \theta_{1},\cdots,\theta_{n}\right)  \rangle_{1\cdots
N}\nonumber\\
&=U_{n}W\left(  \theta_{n}\right)  U_{n-1}W\left(  \theta_{n-1}\right)  \cdots U_{1}W\left(\theta_{1}\right)  |\Psi\rangle_{12\cdots N}.
\end{align}
From the same calculation as in Eq.\eqref{eq_one-parameter}, we get the
following expression:
\begin{align}
& |\Psi\left(  \theta_{1},\cdots,\theta_{n}\right)  \rangle_{1\cdots
N}\nonumber\\
&=
\left(  \frac{1}{\sqrt{d}}\right)  ^{n+1}
{\displaystyle\sum\limits_{w_{1}\cdots w_{n}}}
\exp\left[  i\sum_{j=1}^{n}w_{s_{j}}\theta_{j}\right]
{\displaystyle\sum\limits_{\bar{s}=1}^{d}}
|\bar{s}\rangle_{1}|\phi\left(  s_{1}\cdots s_{n}\bar{s}\right)
\rangle_{2\cdots N}\nonumber\\
&\quad +O\left(  d^{-(N-3)/2}\right) \label{12}
\end{align}
where
\begin{equation}
\langle\phi\left(  s_{1}\cdots s_{n}\bar{s}\right)  |\phi\left(  s_{1}
^{\prime}\cdots s_{n}^{\prime}\bar{s}^{\prime}\right)  \rangle=\delta
_{s_{1}s_{1}^{\prime}}\cdots\delta_{s_{n}s_{n}^{\prime}}\delta_{\bar{s}\bar
{s}^{\prime}}\label{13}
\end{equation}
holds. Eq. (\ref{13}) guarantees
existence of unitary operator $U_{code}$ which is independent of $\left(
\theta_{1},\cdots,\theta_{n}\right)  $ and satisfies
\begin{align}
& |\bar{s}\rangle_{1}|\phi\left(  s_{1}\cdots s_{n}\bar{s}\right)
\rangle_{2\cdots N}\nonumber\\
&=U_{code}|s_{1}\rangle_{1}|s_{2}\rangle_{2}\cdots
|s_{n}\rangle_{n}|\bar{s}\rangle_{n+1}|\chi\rangle_{n+2\cdots N}.
\end{align}
Substituting the above equation into eq. (\ref{12}), our main result of
decoupled QIC state is derived:
\begin{align}
&|\Psi\left(  \theta_{1},\cdots,\theta_{n}\right)  \rangle_{AB}  \nonumber\\
&=U_{code}|\varphi(\theta_{1})\rangle_{1}|\varphi(\theta_{2})\rangle_{2}
\cdots|\varphi(\theta_{n})\rangle_{n}|\varphi(0)\rangle_{n+1}|\chi
\rangle_{n+2\cdots N}\nonumber\\
& +O\left(  d^{-(N-3)/2}\right)  ,\label{16}
\end{align}
where each QIC state is given by
\[
|\varphi(\theta_{j})\rangle=\frac{1}{\sqrt{d}}
{\displaystyle\sum\limits_{s_{j}=1}^{d}}
e^{iw_{s_{j}}\theta_{j}}|s_{j}\rangle.
\]
The quantum state in Eq.\eqref{16} can be obtained by permuted multi-parameter write operations and different random unitary operations:
\begin{align}
& |\Psi\left(  \theta_{1},\cdots,\theta_{n}\right)  \rangle_{1\cdots N}\nonumber\\
&=U_{n}^{\prime}W\left(  \theta_{a_{n}}\right)  U_{n-1}^{\prime
}W\left(  \theta_{a_{n-1}}\right)  \cdots U_{1}^{\prime}W\left(
\theta_{a_{1}}\right)  U_0^{\prime}|0\rangle_{12\cdots N}\nonumber\\
&\quad+O\left(  d^{-(N-3)/2}\right),
\end{align}
where $\left(
a_{1},\cdots,a_{n-1},a_{n}\right)$ is a permutation of $\left(  1,\cdots,n-1,n\right)  $. 
This result is obtained in the case of the write operations to the first qubit. It can be extended to the cases of port-changed write operations. At every step of write operations, the location change of input port is allowed such that
\begin{align}
& |\Psi\left(  \theta_{1},\cdots,\theta_{n}\right)  \rangle_{1\cdots N}\nonumber\\
&=U_{n}^{\prime\prime}W_{p_n}\left(  \theta_{n}\right)  U_{n-1}^{\prime\prime
}W_{p_{n-1}}\left(  \theta_{n-1}\right)  \cdots U_{1}^{\prime\prime}W_{p_1}\left(
\theta_{1}\right)  U_0^{\prime\prime}|0\rangle_{12\cdots N}\nonumber\\
&+O\left(  d^{-(N-3)/2}\right),
\end{align} 
where $W_p(\theta_j)$ is the write operation on the $p$-th qudit:
\begin{align}
 W_p(\theta_j)\equiv I_1\otimes I_{p-1}\otimes e^{i\theta_j\sigma}\otimes I_{p+1}\otimes I_{N}.
\end{align}
Moreover, we get the same result in Eq.\eqref{12} even if we change the generator of the write operations at each step as $ \sigma_j=u_j\sigma u_j^\dag$, where $u_j$ is a single-qudit unitary operator.
The decoupled QIC state in Eq.\eqref{16} gives the rotational isometry in the Fisher information metric.

\begin{figure}[tbp]
\includegraphics[width=6.0cm]{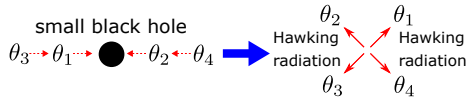}
\caption{Schematic picture for the decoupled QICs in black hole creation.}\label{figure_bh}
\end{figure}

\textit{QIC Speculation of Black Hole Evaporation.}--- An interesting
application of the decoupled QIC state may be small black hole creation in
future particle colliders \cite{SG}. Figure \ref{figure_bh} depicts the setup. 
Suppose that particles with
high energy carry different information $\theta_{j}$. A
small black hole in a high temperature may be created, and decay into the Hawking
radiation. If the black hole temperature is high enough, the information of
each $\theta_{j}$ may be confined in each QIC of the radiation. The model
roughly predicts the emergence of rotational isometry of quantum Fisher
information metric $\left[  g_{jj^{\prime}}\right]  =F\delta_{jj^{\prime}}$. Even though this conjecture of emergent isometry is
merely a bold argument, investigation of quantum Fisher information of
Hawking radiation about unknown multiple parameters is promising.

\textit{Conclusion.}--- We have proven the emergence of decoupled QICs in the
double limits of large $N$ and high temperature as shown in eq. (\ref{16}).
The results may be applied to small black hole creation and roughly predicts the
rotational isometry of quantum Fisher information metric. The conjecture may
be tested in future experiments of cold atoms in an entangled state under fast scrambling.
\begin{acknowledgments}
\textit{Acknowledgments}.---
We would like to thank H. Tajima and S. Endo for useful discussions about
quantum Fisher information detection. 
We also would like to thank the participants in Relativistic Quantum Information-North 2019 for valuable comments.
This research is partially supported by
JSPS KAKENHI Grant Number JP19K03838 (M.H.) and JP18J20057 (K.Y.), and by Graduate Program on Physics for the Universe (GP-PU), Tohoku University (K.Y.).
\end{acknowledgments}

\clearpage
\widetext

\section{Supplemental material}
\setcounter{equation}{0}
\makeatletter
\renewcommand{\theequation}{    S\arabic{equation}} 
\@addtoreset{equation}{section} \makeatother
\begin{center}
{\large Ensemble formula }
\end{center}
Here, we derive formulas used in this Letter to calculate the ensemble average over the unitary matrices distributed according to the Haar measure. 

Let $U$ be a unitary operator on a Hilbert space $\mathcal{H}$. Introducing an orthonormal basis $\left\{|\Psi_i\rangle\right\}_{i=1}^{D}$ with $D\equiv \mathrm{dim}\mathcal{H}$, the unitary matrix is expanded as $U=\sum_{ij=1}^DU_{ij}|\Psi_i\rangle\langle\Psi_j|$, where $U_{i,j}\equiv \langle\Psi_i|U|\Psi_j\rangle$. Since the Haar measure is both left and right invariant, its average takes the following form:
\begin{align}
 \overline{U_{i,j}U_{k,l}^*}&= a \delta_{il}\delta_{jk}\label{eq_sm_ave1}
\end{align}
with some number $a$, where the over-line denotes the ensemble average over the Haar measure. From the normalization condition $\overline{1}=1$, we get
\begin{align}
 \sum_{j=1}^{D} \overline{U_{i,j}U_{k,j}^*}&=\delta_{ik}.
\end{align}
On the other hand, from Eq.~\eqref{eq_sm_ave1}, 
\begin{align}
 \sum_{j=1}^{D}\overline{U_{i,j}U_{k,j}^*}=a\sum_{j=1}^D\delta_{ij}\delta_{jk}=aD\delta_{ik}
\end{align}
holds, implying that
\begin{align}
\overline{U_{i,j}U_{k,l}^*}&= \frac{1}{D} \delta_{il}\delta_{jk}.
\end{align}

Repeating the above arguments, we can derive the ensemble average formula for higher moments. From the symmetry, it holds that
\begin{align}
\overline{U_{i,j}U_{k,l}U_{x,y}^*U_{z,w}^*}=b\left(\delta_{iy}\delta_{jx}\delta_{kw}\delta_{lz}+\delta_{iw}\delta_{jz}\delta_{ky}\delta_{lx}\right)+c\left(\delta_{iy}\delta_{jz}\delta_{kw}\delta_{lx}+\delta_{iw}\delta_{jx}\delta_{ky}\delta_{lz}\right)\label{eq_sm_ave2}
\end{align}
for numbers $b,c$. The coefficients are determined from constraints
\begin{align}
\sum_{i,j,k=1}^D\overline{U_{i,j}U_{k,l}U_{j,i}^*U_{z,k}^*}=D\delta_{lz}\label{eq_sm_ave3}
\end{align}
and
\begin{align}
 \sum_{i,k,l=1}^D\overline{ U_{i,j}U_{k,l}U_{l,i}^*U_{z,k}^*}=\delta_{jz}.\label{eq_sm_ave4}
\end{align}
Combining eqs.~\eqref{eq_sm_ave2},\eqref{eq_sm_ave3},\eqref{eq_sm_ave4}, we get
\begin{align}
 \overline{ U_{i,j}U_{k,l}U_{x,y}^*U_{z,w}^*}=\frac{1}{D^2-1}\left(\delta_{iy}\delta_{jx}\delta_{kw}\delta_{lz}+\delta_{iw}\delta_{jz}\delta_{ky}\delta_{lx}\right)-\frac{1}{D(D^2-1)}\left(\delta_{iy}\delta_{jz}\delta_{kw}\delta_{lx}+\delta_{iw}\delta_{jx}\delta_{ky}\delta_{lz}\right).
\end{align}

\begin{center}
{\large Canonical typicality}
\end{center}
For macroscopic systems, it is known that an overwhelming majority of pure states are almost indistinguishable locally \cite{L,SJ,SA,S,S1}. We review this result here, based on the arguments in \cite{S,S1}. Define a set of traceless Hermitian operators $\left\{\lambda_i\right\}_{i=1}^{D^2-1}$ on a Hilbert space $\mathcal{H}_s$ satisfying $\mathrm{Tr}_{\mathcal{H}_s}\left(\lambda_i\lambda_j\right)=d_s\delta_{ij}$, where $d_s$ is the dimension of Hilbert space: $d_s\equiv\dim{\mathcal{H}_s}$. To investigate a macroscopic system, let us consider $N(\gg 1)$ copies of the system. A basis of the traceless Hermitian operators on $\mathcal{H}=\mathcal{H}_s^{\otimes N}$ can be constructed by $\left\{\lambda_{\mu}\right\}_{\mu}$, where we have defined
\begin{align}
  \lambda_{\mu}\equiv\lambda_{\mu_1}\otimes\lambda_{\mu_2}\otimes\cdots\otimes\lambda_{\mu_N}
 =\bigotimes_{n=1}^N \lambda_{\mu_{n}} \qquad
 (\mu=(\mu_1,\cdots,\mu_N), \,\mu_n=0,\cdots d_s^2-1)
\end{align}
and $\lambda_0\equiv\mathbb{I}$.
Since this basis is orthonormal
\begin{align}
 \mathrm{Tr}_{\mathcal{H}}\left(\lambda_{\mu}\lambda_{\nu}\right)=D\delta_{\mu\nu},\quad D\equiv d^N ,
\end{align}
any density operator $\rho$ can be expanded as 
\begin{align}
 \rho=\frac{1}{D}\sum_{\mu=0}^{D^2-1}\langle{\lambda_{\mu}}\rangle\lambda_{\mu},
\end{align}
where we have defined $\langle\lambda_{\mu}\rangle\equiv\mathrm{Tr}_{\mathcal{H}}\left(\rho\lambda_{\mu}\right)$. For macroscopic systems, it is practically impossible to perform a measurements on the whole $N$-body system. Therefore, the reduced state for $m(\ll N)$-body system plays an important role. As an example, consider the reduced state for the subsystem $A$ composed of the first $m$ body. The reduced state for the subsystem $A$ is given by
\begin{align}
 \rho_A=\frac{1}{d_s^m}\sum_{\mu_1=0}^{d^2-1} \cdots
 \sum_{\mu_m=0}^{d^2-1}\mathrm{Tr}_{\mathcal{H}^{\otimes N}}\left(\rho\lambda_{(\mu_1,\cdots,\mu_m,0,\cdots,0)}\right)\lambda_{(\mu_1,\cdots,\mu_m)}.
\end{align}
Let $H$ be the Hamiltonian of the system. By using the energy eigenvectors $H|E_i\rangle=E_i|E_i\rangle$, the microcanonical ensemble asserts that the system is in a pure state randomly chosen from an energy shell $[E-\delta E,E]$ ($\delta E/E\ll1$). To show the typicality of the reduced state, let us calculate the expectation value and the variance a Haar-random pure state distributed over $\mathcal{H}_{[E-\delta E,E]}$, where $\mathcal{H}_{[E-\delta E,E]}$ is the Hilbert space spanned by energy eigenstates $E_i$ satisfying $E_i\in[E-\delta E,E]$. A Haar-random pure state is generated by $|\Psi(U)\rangle\equiv U|\Psi_0\rangle$, where $|\Psi_0\rangle\in\mathcal{H}_{[E-\delta E,E]}$ is a fixed pure state and $U$ is a unitary operator on $\mathcal{H}_{[E-\delta E,E]}$ randomly chosen according to the Haar measure. For any Hermite operator $\lambda$, the ensemble average is calculated by using Eq.~\eqref{eq_sm_ave1} as
\begin{align}
\overline{\langle\Psi(U)|\lambda|\Psi(U)\rangle}=\frac{1}{d_E}\sum_{i=1}^{d_E}\langle E_i|\lambda|E_i\rangle,\label{eq_sm_aveobs1}
\end{align}
where we have defined $d_E\equiv \mathcal{H}_{[E-\delta E,E]}$ and $\left\{|E_i\rangle\right\}_{i=1}^{d_E}$ are energy eigenstates in $\mathcal{H}_{[E-\delta E,E]}$. Similarly, from eq.~\eqref{eq_sm_ave4}, we get
\begin{align}
 \overline{\langle\Psi(U)|\lambda|\Psi(U)\rangle ^2}=\frac{1}{d_E(d_E+1)}\left(\sum_{ij=1}^{d_E}\langle E_{j}|\lambda|E_i\rangle\langle E_i|\lambda|E_{j}\rangle+\sum_{i,j=1}^{d_E}\langle E_i|\lambda|E_i\rangle\langle E_j|\lambda|E_j\rangle\right).\label{eq_sm_aveobs2}
\end{align}
Combining eqs.~\eqref{eq_sm_aveobs1} and \eqref{eq_sm_aveobs2}, we obtain the upper-bound for the variance as follows:
\begin{align}
& \overline{\langle \Psi(U)|\left(\lambda-\langle\Psi(U)|\lambda|\Psi(U)\rangle\right)^2|\Psi(U)\rangle}\nonumber\\
&=\frac{1}{d_E(d_E+1)}\sum_{i,j=1}^{d_E}\left|\langle E_i|\lambda|E_j\rangle\right|^2-\frac{1}{d_E^2(d_E+1)}\left(\sum_{i=1}^{d_E}\langle E_i|\lambda|E_i\rangle\right)^2\nonumber\\
 &\leq  \frac{1}{d_E(d_E+1)}\sum_{i,j=1}^{d_E}\left|\langle E_i|\lambda|E_j\rangle\right|^2\leq \frac{1}{d_E(d_E+1)}\sum_{i}^{d_E}\left|\langle E_i|\lambda^2|E_i\rangle\right|\leq \frac{1}{d_E+1}\left|\lambda^2\right|,\label{eq_sm_ub1}
\end{align}
where $\left|\lambda^2\right|$ denotes the maximum eigenvalue of $\lambda^2$. For a fixed $U$, the reduced state for a $m$-body system is calculated as
\begin{align}
 \rho(U)=\frac{1}{d_s^m}\sum_{\mu_1=0}^{d^2-1} \cdots
 \sum_{\mu_m=0}^{d^2-1}\langle\Psi(U)|\lambda_{(\mu_1,\cdots,\mu_m,0,\cdots,0)}|\Psi(U)\rangle\lambda_{(\mu_1,\cdots,\mu_m)}
\end{align}
The averaged state is given by 
\begin{align}
 \rho_A^{\text{(ave.)}}=\mathrm{Tr}_{\mathcal{H}_{\bar{A}}}\left(\frac{1}{d_E}\sum_{i=1}^{d_E}|E_i\rangle\langle E_i|\right),
\end{align}
where $\mathcal{H}_{\bar{A}}$ denotes the Hilbert space for the complement system of the subsystem $A$. By using the bound in eq.~\eqref{eq_sm_ub1}, we get
\begin{align}
\overline{\mathrm{Tr}_{\mathcal{H}_A}\left(\left(\rho_A(U)-\rho_A^{\text{(ave.)}}\right)^2\right)}\leq\frac{d_s^{2m}}{d_E+1},
\end{align}
where we have used the fact that $\left|\lambda_{(\mu_1,\cdots,\mu_m,0,\cdots,0)}^2\right|\leq d_s^m$. For thermodynamically normal system, $d_E$ grows exponentially as $N$ increase. Therefore, for large $N$, the reduced state for $m$-body system is well approximated by $\rho_A^{\text{(ave.)}}$. Now, suppose that the Hamiltonian of the system is given by $H=H_A\otimes \mathbb{I}_{\bar{A}}+\mathbb{I}_A\otimes H_{\bar{A}}$. Then, the energy eigenstate is constructed as $\left\{|E_A\rangle\otimes |E_{\bar{A}\rangle}\right\}_{E_A,E_{\bar{A}}}$, where $|E_A\rangle$ and $|E_{\bar{A}}\rangle$ are eigenstates for $H_A$ and $H_{\bar{A}}$, respectively. By using the density of states $\Omega_{\bar{A}}(E_{\bar{A}})$ for the subsystem $\bar{A}$, the average reduced state is given by
\begin{align}
 \rho_A^{\text{(ave.)}}&=\mathrm{Tr}_{\mathcal{H}_{\bar{A}}}\left(\frac{1}{d_E}\sum_{E_A,E_{\bar{A}};E_A+E_{\bar{A}}\in[E-\Delta,E]}|E_A\rangle\langle E_A|\otimes |E_{\bar{A}}\rangle\langle E_{\bar{A}}|\right)\nonumber\\
 &\approx\frac{1}{d_E}\sum_{E_A}\Omega_{\bar{A}}(E-E_A)|E_A\rangle\langle E_A| \approx \frac{1}{Z_\beta}\sum_{E_A}e^{-\beta E_A}|E_A\rangle\langle E_A|,
\end{align}
where $\beta\equiv\frac{\partial}{\partial E}\ln{\Omega_{\bar{A}}(E)}$ is the inverse temperature and $Z_\beta\equiv \sum_{E_A}e^{-\beta E_A}$ is the partition function. Therefore, the reduced state for a small subsystem A of an overwhelming majority of overall pure states distributed uniformly over an energy shell is approximately given by a Gibbs state.


\begin{thebibliography}{99}                                                 
\bibitem {HP}P. Hayden and J. Preskill, JHEP 09, 120 (2007).

\bibitem {SS}Y. Sekino and L. Susskind, JHEP 10, 065 (2008).

\bibitem {BP}S. L. Braunstein and A. K. Pati, Phys. Rev. Lett. 98:080502 (2007).

\bibitem {P}K. Modi, A. K. Pati, A. Sen, and U. Sen, Phys. Rev. Lett. 120,
230501 (2018).

\bibitem {ED}R. Islam, R. Ma, P. M. Preiss, M. E. Tai, A. Lukin, M. Rispoli,
and M. Greiner, Nature (London) 528, 77 (2015).

\bibitem {SU}T. Shitara and M. Ueda, Phys. Rev. A 94, 062316 (2016).

\bibitem{RD} R. Blatt and D. Wineland, Nature 453, 1008 (2008).

\bibitem{SKA} S. Takeda, K. Takase and A. Furusawa, Science Advances 5, 5 (2019).

\bibitem {TYH2}J. Trevison, K. Yamaguchi, and M. Hotta, Prog. Theo. Exp. Phys.
10, 103A03 (2018).

\bibitem {TYH}J. Trevison, K. Yamaguchi, and M. Hotta, J. Phys. A, Math and
Theo. 52, 125402 (2019).

\bibitem {YWH}K. Yamaguchi, N. Watamura, and M. Hotta, Phys. Lett. A 383, 1255 (2019).

\bibitem {L}S. Lloyd, Ph.D. Thesis, Rockefeller University (1988).

\bibitem{SJ} S. Goldstein, J. L. Lebowitz, R. Tumulka, and N. Zangh\`{i}, Phys. Rev.Lett. 96, 050403 (2006).

\bibitem{SA}S. Popescu, A. J. Short, and A. Winter, Nat. Phys. 2, 754 (2006).

\bibitem {S}A. Sugita, RIMS Kokyuroku (Kyoto) 1507, 147 (2006).

\bibitem {S1}A. Sugita, Nonlinear Phenom. Complex Syst. 10, 192 (2007).

\bibitem{SM} See Supplemental Material for more detailed derivations.

\bibitem{SBG1} D. M. Stamper-Kurn, M. R. Andrews, A. P. Chikkatur, S. Inouye, H.-J. Miesner, J. Stenger, and W. Ketterle, Phys. Rev. Lett. 80, 2027 (1998).

\bibitem{SBG2} Dan M. Stamper-Kurn and Masahito Ueda, Rev. Mod. Phys. 85, 1191 (2013).

\bibitem{FG1} S. Taie, R. Yamazaki, S. Sugawa, and Y. Takahashi, Nat. Phys. 8, 825 (2012).

\bibitem{FG2} H. Ozawa, S. Taie, Y. Takasu, and Y. Takahashi, Phys. Rev. Lett. 121, 225303 (2018).

\bibitem {Lubkin}E. Lubkin, J. Math. Phys. 19, 1028 (1978).

\bibitem {Seth_Pagels}S. Lloyd and H. Pagels, Ann. Phys. 188, 186 (1988).

\bibitem{Page_sub} D. N. Page, Phys. Rev. Lett. 71, 1291 (1993).

\bibitem{SG}S. Dimopoulos and G. Landsberg, Phys. Rev. Lett. 87, 161602  (2001).

\end{thebibliography}
\end{document}